\documentclass[aps,pra,twocolumn,superscriptaddress]{revtex4-1}

\usepackage{graphicx}
\usepackage{amssymb}
\usepackage{subfigure}
\usepackage{amsmath}
\usepackage{amsfonts}
\usepackage{bm}
\usepackage{color}
\usepackage{palatino}
\newcommand{\ket}[1]{\ensuremath{\left|{#1}\right\rangle}}

\begin{document}

\title{Experimental investigation of linear-optics-based quantum target detection}

\author{G. H. Aguilar}
\affiliation{Instituto de F\'{i}sica, Universidade Federal do Rio de Janeiro, Caixa Postal 68528, Rio de Janeiro, RJ 21941-972, Brazil}
\author{M. A. de Souza}
\affiliation{Instituto de F\'{i}sica, Universidade Federal do Rio de Janeiro, Caixa Postal 68528, Rio de Janeiro, RJ 21941-972, Brazil}
\affiliation{ Instituto Nacional de Metrologia,  Qualidade e Tecnolog\'ia Industrial (INMETRO), Duque de Caxias, RJ 25250-020, Brazil}
\author{R. M. Gomes}
\affiliation{Instituto de F\'{i}sica, Universidade Federal de Goi\'{a}s, 74.001-970, Goi\^{a}nia, Goi\'{a}s, Brazil}
\author{J. Thompson}
\affiliation{Centre for Quantum Technologies, National University of Singapore, 3 Science Drive 2, 117543, Singapore}
\author{M. Gu}
\affiliation{Centre for Quantum Technologies, National University of Singapore, 3 Science Drive 2, 117543, Singapore}
\affiliation{School of Physical and Mathematical Sciences, Nanyang Technological University, Singapore 639673, Singapore}
\affiliation{Complexity Institute, Nanyang Technological University, Singapore 639673, Singapore}
\author{L. C. C\'{e}leri}
\affiliation{Instituto de F\'{i}sica, Universidade Federal de Goi\'{a}s, 74.001-970, Goi\^{a}nia, Goi\'{a}s, Brazil}
\author{S. P. Walborn}
\affiliation{Instituto de F\'{i}sica, Universidade Federal do Rio de Janeiro, Caixa Postal 68528, Rio de Janeiro, RJ 21941-972, Brazil}

\begin{abstract} 
	The development of new techniques to improve measurements is crucial for all sciences. By employing quantum systems as sensors to probe some physical property of interest allows the application of quantum resources, such as coherent superpositions and quantum correlations, to increase measurement precision. Here we experimentally investigate a scheme for quantum target detection based on linear optical measurment devices, when the object is immersed in unpolarized background light.  By comparing the quantum (polarization-entangled photon pairs) and the classical (separable polarization states), we found that the quantum strategy provides us an improvement over the classical one in our experiment when the signal to noise ratio is greater than 1/40, or about 16dB of noise. This is in constrast to quantum target detection considering non-linear optical detection schemes, which have shown resilience to extreme amounts of noise.  A theoretical model is developed which shows that, in this linear-optics context, the quantum strategy suffers from the contribution of multiple background photons.  This effect does not appear in our classical scheme.  By improving the two-photon detection electronics, it should be possible to achieve a polarization-based quantum advantage for a signal to noise ratio that is close to 1/400 for current technology.  \end{abstract}

\maketitle

\noindent

\section{Introduction} 
Quantum target detection (QTD) is a promising area of quantum technologies that harnesses uniquely quantum effects such as entanglement and coherent superpositions, to improve sensing of physical quantities  \cite{Degen, Pirandola18, Genovese16}. In particular it has been established that quantum correlations can improve the ability to resolve a parameter of a physical system, even in environments which sustain moderate levels of noise. This is exemplified by Quantum Illumination protocols which harness quantum entanglement to improve our ability to resolve a faintly reflective target bathed in intense environmental noise \cite{lloyd08,Tan08, Barzanjeh15, Weedbrook,Bradshaw17}. Experiments using multimode gaussian light and a measurement scheme based on an optical parametric amplifier demonstrated the robustness of QTD in the high loss and high noise regime \cite{Zhang15}. It was recently shown that a measurement device based on sum frequency generation and feed-forward can improve these previous results \cite{Zhang17}.  
\par 
In the original QTD proposal by Lloyd \cite{lloyd08}, a pair of photons in a $d \times d$ dimensional entangled state is used as the source.  One of the photons is sent to probe the object, and a joint measurement is performed on the photon pair.  A simple measurement based on two-photon coincidence detection has been demonstrated to exploit non-classical spatial and temporal correlations in order to improve signal to noise ratio beyond classical light benchmarks \cite{Lopaeva13,Ragy14,England18}.   However, as shown in \cite{lloyd08}, the optimal  measurement is one that separates the state produced by the source from the orthogonal subspace.  In general, this measurement requires a non-linear optical medium \cite{lloyd08,Lamata07}, which typically suffers from low efficiency \cite{kim01}, rendering the protocol highly probabilistic.  However, in the simplest case of $2 \times 2$ dimensional systems, it is well known that projection onto a single maximally entangled Bell state is deterministically possible  with linear-optical elements alone \cite{mattle96,walborn03b}.  Motivated by this fact, we implement and study a target detection protocol using  photon pairs in a linear-optics setup. One photon is sent to probe the presence of a target (reflecting object) that is immersed in unpolarized background light (noise environment). Both ``classical" and quantum strategies are employed. The former employs photons in a separable polarization state, while the quantum one uses polarization entanglement between the photons. In both strategies, the goal is to obtain information regarding the presence or absence of the target.  Our results show that polarization entanglement provides an enhancement in our ability to identify the object when the signal to noise ratio is $\gtrsim 1/40$. This is due to the fact that in the linear-optics regime, the quantum strategy is limited by multi-photon contributions from the noise source.  In order to completely describe the experimental results, we developed a theoretical model for our protocol by considering the relevant detection events that are caused by the background noise, obtaining very good agreement between theory and experiment.  We show how the linear optical scheme can be improved using current technology.  Our results shows that to exploit polarization entanglement in the very high noise regime, one must employ non-linear optical devices.       

\vspace{0.2cm} 
\noindent 
\section{Target detection with linear optics} 
\begin{figure}
	\includegraphics[width=7cm]{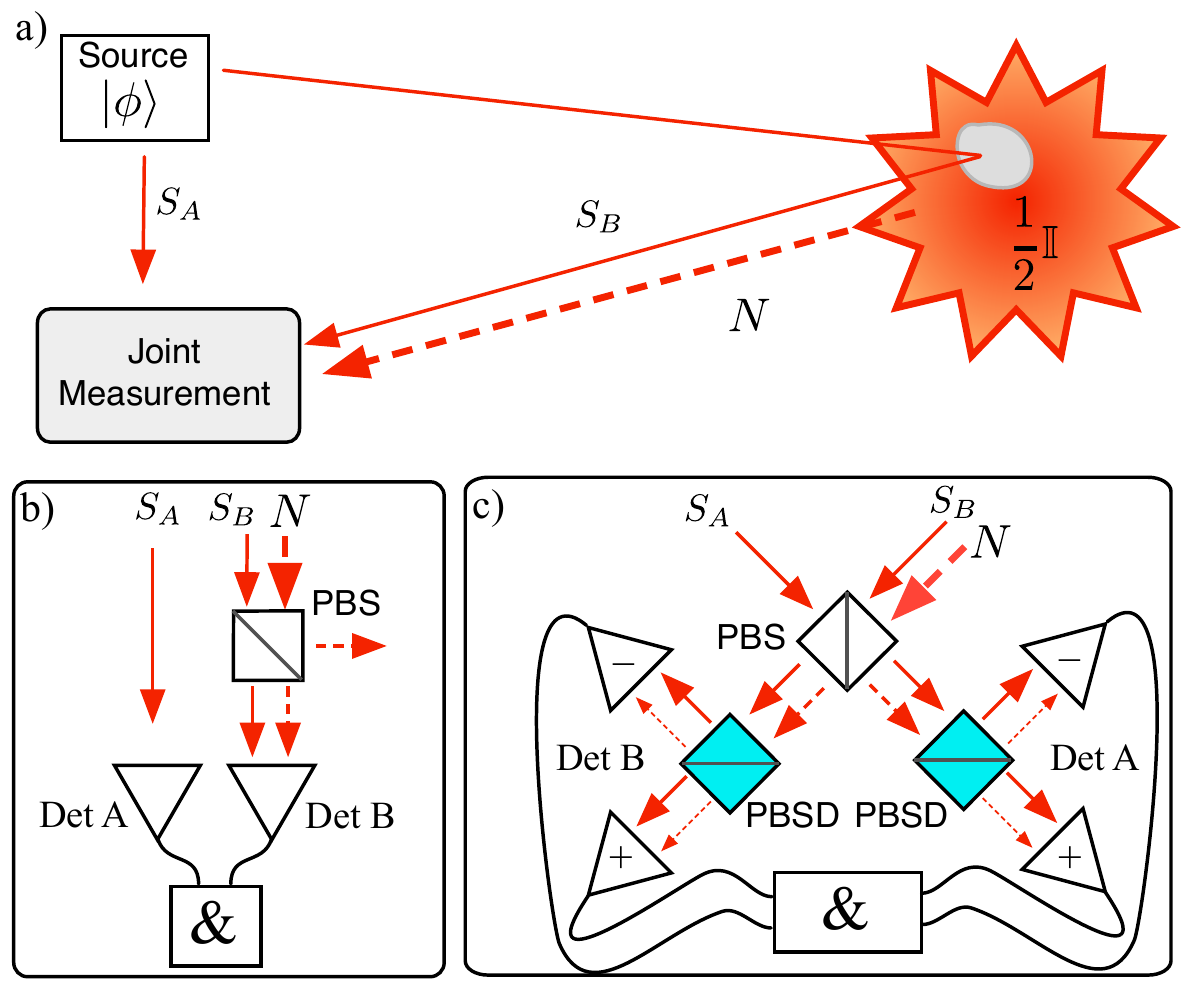}
	\caption{Quantum target detection scheme using photon pairs to probe for the presence of an object immersed in unpolarized background radiation. Panel a) illustrates the case when the object is present, both the signal photons $S_B$ and the unpolarized background photons $N$ return to the joint measurement stage. When the object is absent (not illustrated), the signal photon $S_B$ does not return for joint measurement. Panel b) shows the joint polarization measurement scheme using a local projective measurement when the initial state $\ket{\phi}$ is a product state, while panel c) considers the projection onto a Bell-state using linear optics when $\ket{\phi}$ is an entangled state. PBS is a polarizing beam splitter separating horizontal and vertical polarizations, and PBSD is a polarizing beam splitter in the linear diagonal polarization basis.  In both the classical and quantum strategies considered here two-photon coincidence detection is employed.}
	\label{fig:plot}
\end{figure}

Inspired by Refs. \cite{lloyd08, Weedbrook}, let us consider a particular target detection scheme, in which an object, which may or not be present, is immersed in a background of unpolarized light, as illustrated in Fig. \ref{fig:plot} a).  Photons from a light source are reflected by the object to a detector, so that detection of a signal photon indicates that the object is present. The main goal is to distinguish between the signal photons that are reflected from the object from the noise photons originating from the background radiation. Here we consider only the polarization degree of freedom of the light source and background, so that the photons can be treated as qubits. We further assume that the source emits pairs of photons in some bipartite state $\ket{\phi}_{AB}$, and that the signal photon $B$, is sent to probe the object, while photon $A$ is isolated, and later used only in the final joint measurement stage.

We consider a joint measurement using only linear optics devices, with the goal of distinguishing between the two situations described above, in order to best detect the presence or absence of the object. Below we will assume the general joint measurement strategy of projecting onto the initial state $\ket{\phi}$ or the orthogonal subspace. Let us define the probability to obtain result $r$ conditioned on the presence of the object as $p(r|x)$, where $r=0$ indicates projection onto state $\ket{\phi}$ while $r=1$ corresponds to projection onto the orthogonal subspace. $X = \lbrace x,p(x)\rbrace$ is a binary random variable indicating the presence ($x=0$) or absence ($x=1$) of the object. If there are no noise or imperfections, we expect $p(0|0)=1$ and $p(1|0)=0$ for any strategy. In what follows we define the classical and quantum strategies for target detection.

\subsection{Classical Strategy: Product state}
 The classical strategy for quantum target detection here is to prepare a pair of photons in a product state of their polarization, $\ket{\phi} = \ket{\alpha}_A \ket{\beta}_B$, with $\alpha$ and $\beta$ labeling the polarization. In the present context considering only the polarization degree of freedom, this scheme is similar to preparing a single probe photon, sending it to interact with the target object, and performing a local polarization measurement.  However, coincidence detection of photon pairs can offer additional advantages over a classical laser source due to the temporal correlations, as has been explored in Ref. \cite{England18}.  For this reason, we consider here a pair of photons in a product state, as it allows us isolate the role of polarization entanglement, and provide a fair comparison between the classical and quantum strategy arising from the polarization degree of freedom alone.  We note that via the coincidence detection both our classical and quantum schemes offer improvements when compared to a classical laser source, and that the enhancement due to polarization entanglement appears in addition to the that due to the temporal correlations.  
 \par
 If we assume a local polarization projection in mode $B$, as illustrated in \ref{fig:plot} b), one can eliminate half of the unpolarized noise by projecting onto the polarization state $\ket{\beta}$. Using the subscript $c$ for variables corresponding to the classical case, let us define the conditional probability as $p_c(r|x)$. We note that, for any amount of unpolarized noise, $p_c(r|1)=1/2$, when the object is absent.

\subsection{Quantum Strategy: Polarization entanglement and linear-optics Bell-state projection}

 Our quantum strategy consists in preparing an initial state given by the entangled Bell state $\ket{\phi}=\ket{\phi^+}$, where we define
\begin{equation}
 \ket{\phi^\pm}  = \frac{1}{\sqrt{2}}(\ket{HH}_{AB} \pm  \ket{VV}_{AB}), 
\end{equation}
 and $H$ ($V$) refers to horizontal (vertical) polarization. 
  We will use the subscript $q$ to denote variables relevant to the quantum case, and thus refer to the conditional probabilities here as $p_{q}(r|x)$. The noise as well as photon $A$ (alone) are completely unpolarized, when the object is absent we expect $p_{q}(0|1)=1/4$ and $p_{q}(1|1)=3/4$ for any amount of noise.  This follows from the fact that a completely unpolarized bipartite state can be written as a convex sum of the four Bell states, each with probability 1/4. 

For the detection system we consider here a Bell-state projection using only linear optical elements, which can be performed using two-photon Hong-Ou-Mandel interference \cite{hom87,mattle96,walborn03b}. Here we choose a partial Bell-state analyzer (BSA) based on three polarizing beam splitters (PBSs) \cite{Aguilar12}, as shown in Fig. \ref{fig:plot} c), though other schemes are possible.  Photons coming from modes $A$ and $B$ are first superposed onto a central PBS that transmits $H$ polarization while reflecting $V$ polarization. Photon pairs in state $\ket{\phi^{\pm}}$ always result in one photon in each of the output ports of the central PBS. We use one half-wave plate and an additional PBS (which we represent together in the figure as PBSD) in each output of the central PBS to separate the diagonal polarization components $\ket{\pm} \equiv (\ket{H} \pm \ket{V})/\sqrt{2}$.  With this scheme, one can then identify the $\ket{\phi^+}$ state by registering  joint detection events at detectors $A_+$ and $B+$, or $A_-$ and $B_-$.   To see this, let us denote the corresponding detection operators as $\hat{d}_{A\pm}$ and $\hat{d}_{B\pm}$, and write the joint detection operators in terms of the operators of the input modes as
\begin{subequations}
\label{eq:ops}
\begin{equation}
	\hat{d}_{A+}\hat{d}_{B+} = \frac{1}{2} (\hat{a}_{H}\hat{b}_{H}+ \hat{a}_{V}\hat{b}_{V} +\hat{a}_{H}\hat{a}_{V} + \hat{b}_{H}\hat{b}_{V}) 
\end{equation}
and
\begin{equation}
	\hat{d}_{A-}\hat{d}_{B- } = \frac{1}{2} (\hat{a}_{H}\hat{b}_{H}+ \hat{a}_{V}\hat{b}_{V} -\hat{a}_{H}\hat{a}_{V} -  \hat{b}_{H}\hat{b}_{V}), 
\end{equation}
\end{subequations}
where operators $\hat{a}$ and $\hat{b}$ refer to the two input spatial modes (polarization can be either $H$ or $V$). In both equations, the first two terms correspond to events that register the input state $\ket{\phi^+}$ . A similar calculation using operators $\hat{d}_{A+}\hat{d}_{B-}$ or $\hat{d}_{A-}\hat{d}_{B+}$ shows that input state $\ket{\phi^+}$ never triggers these detection events.   Moreover, one can show that all other joint detection events correspond to one or more of the other three Bell states.
\par
Let us focus now on the additional terms in Eq. \eqref{eq:ops}.  The third and fourth terms refer to events in which both detected photons come from the same input mode, but with different polarizations.  In the present case this occurs only for the unpolarized background light. Since the unpolarized background is present only in input mode $B$, we can expect a large contribution from the fourth term when the background intensity is sufficiently large, so that there is a significant probability to find more than one background photon in mode $B$.  In other words, there are unwanted joint detection events that result from the joint measurement of two noise photons.  
This does not take place in the classical strategy using coincidence detection, since all of the unpolarized background is routed to a single detector and thus does not contribute to the coincidence counts.  Thus, in contrast to idealized QTD, which might rely on non-linear optical processes for optimal detection \cite{lloyd08,Weedbrook}, we expect the QTD protocol with linear optics to present a quantum advantage only for low intensity noise background.     

\section{Experiment and results} 
\begin{figure}
	\includegraphics[width=7cm]{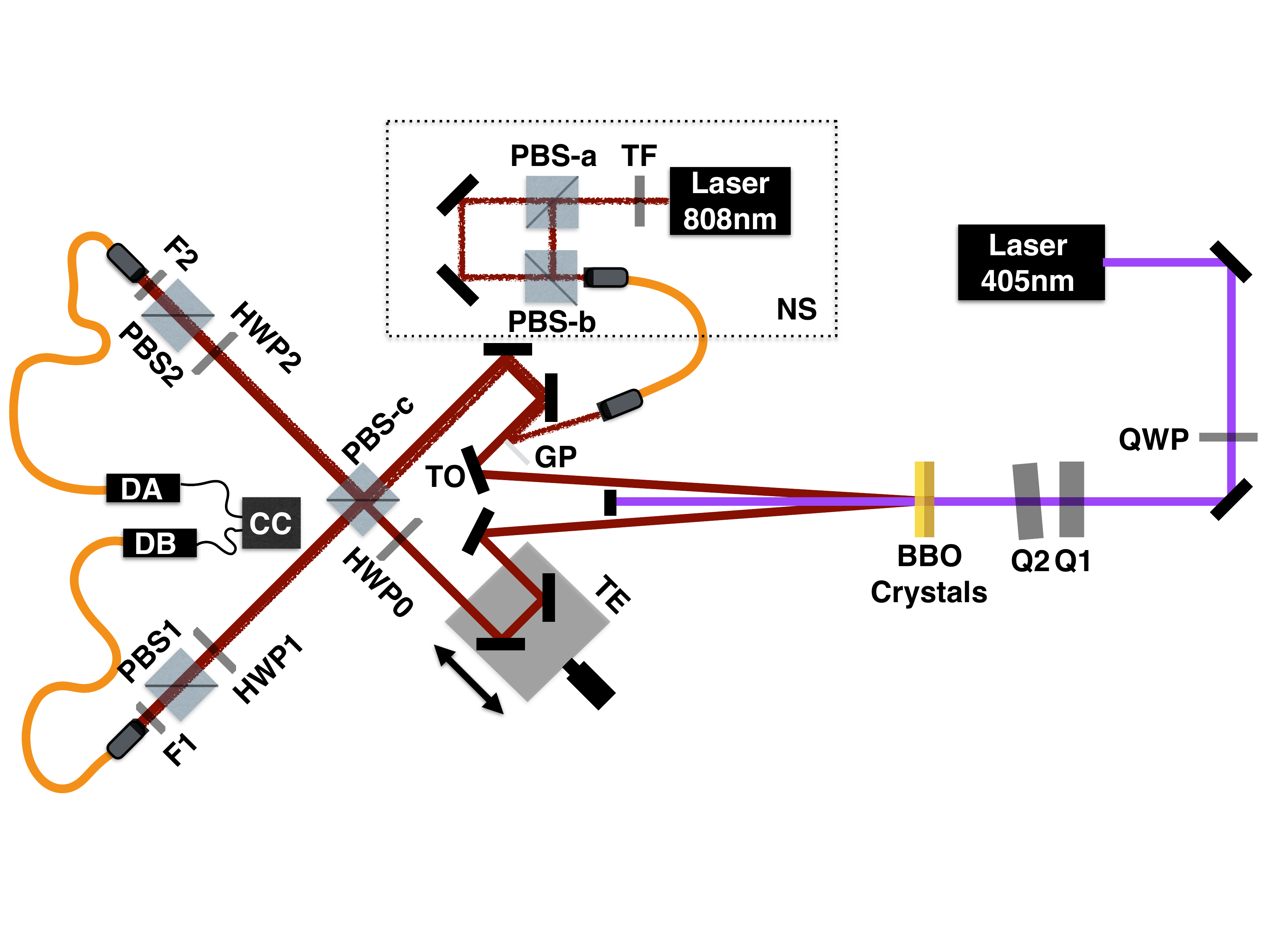}
	\caption{Experimental Setup: A continuous-wave laser at 405 nm pumps two BBO crystals and pairs of photons are produced via parametric down conversion Ref. \cite{Kwiat99}. To compensate polarization walk-off in the source, two birefringent quartz crystals with 5 mm length (Q1 and Q2) are used in the pump beam to finely tune the relative optical delay between the vertical and horizontal polarizations, so that the photon pairs are produced in the state $\ket{\phi^+}$. Photons A and B are sent to a Bell state analyzer, consisting of three PBSs and HWPs (see text).  The path length difference is adjusted to zero using the translation stage TE.  Photon B reflects from the target object TO (a mirror). The unpolarized noise source NS incoherently combines $H$ and $V$ polarized photons from an attenuated diode laser, and these are coupled into mode B using a glass plate GP as a low reflectivity beam splitter. The intensity of the background noise is controlled with a transmission filter TF. Photons  are detected with single photons detectors, and FPGA-based electronics are used to register coincidence counts and single photon events.}
	\label{fig:setup}
\end{figure}

We performed an experimental investigation of the protocols described above using polarization-entangled photon pairs produced from spontaneous parametric down-conversion \cite{Kwiat99}. Our experimental setup is shown in Fig. \ref{fig:setup}.  The target object to be identified is a mirror, marked as TO in the figure. The noise source (NS) is a depolarized laser beam, injected into the signal path using a thin glass plate (GP). We experimentally investigate the QTD protocol for different intensities of the depolarized background noise using the local projective filtering (classical strategy) and linear-optics Bell state projective filtering (quantum strategy) described above. For the quantum case, we create pairs of polarization-entangled photons in the state $\ket{\phi} = \ket{\phi^+}$ \cite{Kwiat99}. Projection onto the Bell-state basis is performed, as described above. For the classical case we produced the initial product state $\ket{\phi} = \ket{HH}$, and removed PBS-c, thus performing only local polarization measurements.  

The joint measurement for both cases ---separable polarization state (classical) and entangled polarization state (quantum)--- involves projection onto the initial polarization state and subsequent detection of the two-photons within a small coincidence window ($\Delta t = 5$ ns) using single-photon detectors.  From these joint detection events we obtain the experimental estimates of the conditional probabilities 
\begin{subequations}
\label{eq:pC}
\begin{align}
p_\pi(0|x) & = \frac{C^{(x)}_{\pi \phi}}{C^{(x)}_{\pi \phi}+C^{(x)}_{\pi \perp}} \\
p_\pi(1|x) & =  \frac{C^{(x)}_{\pi \perp}}{C^{(x)}_{\pi \phi}+C^{(x)}_{\pi \perp}},
\end{align}
\end{subequations}
where $\pi=c,q$ refers to the classical or quantum protocol, $C^{(x)}_{\pi \phi}$ are the coincidence counts corresponding to projection onto initial state $\ket{\phi}$, and $C^{(x)}_{\pi \perp}$ are the counts corresponding to projection onto the subspace orthogonal to $\ket{\phi}$.   

\begin{figure}
	\includegraphics[width=8.8cm]{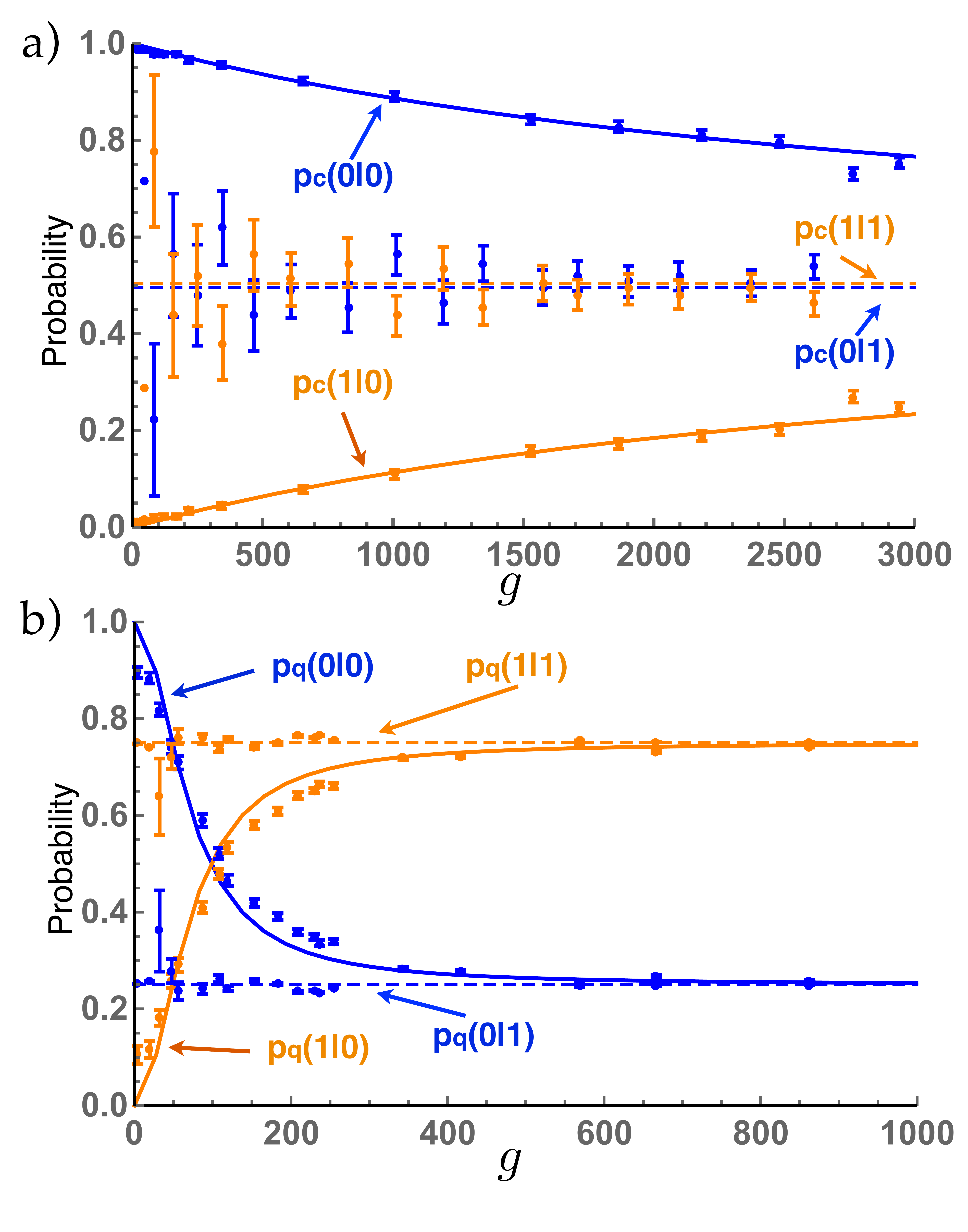}
	\caption{Conditional probabilities. a) Data points corresponding to the measured probabilities $p_c(0|0)$ and $p_c(1|0)$ for the classical strategy as a function of $g$. b) Data points corresponding to the probabilities $p_{bs}(0|0)$ and $p_{bs}(1|0)$ for the quantum strategy with a Bell state analyzer constructed from linear-optical devices. In both plots the dashed lines are theoretical predictions, and the solid curves are obtained from the noise model presented in the main text.} 
	\label{experimental_probs}
\end{figure}

The noise is quantified in terms of the ratio between the count rate of noise photons ($\approx \epsilon N$) versus the count rate of signal photons ($\approx \epsilon S_B$), which we define as $g=N/S_B$, where is $\epsilon$ the efficiency of the detector, and $N$, $S_B$ are the rates of input noise and signal photons, respectively. We note that $g$ can increase due to losses suffered by the signal photons, or due to an increased rate of background photons. The experimental results are shown by the points in Fig. \ref{experimental_probs}.  The dashed lines correspond to the theoretical predictions when the object is absent ($x=1$), which were described in the last section. The solid curves correspond to theoretical curves, which will be developed in the next section.  
\par
For the classical strategy, we remove PBS-c from the setup, sending the idler photons to detector $D_A$ while the signal and noise photons are sent to detector $D_B$ after projection onto the polarization state $\ket{H}$. The experimental results can be observed in Fig. \ref{experimental_probs} a). The absence of the target object TO ($x=1$) was simulated by blocking the path of photon $B$. One can see that the probabilities $p_{c}(0|0) \approx 1$ and $p_{c}(1|0) \approx 0$ for low noise, and tend towards $p_{c}(0|0) \longrightarrow p_{c}(0|1)=1/2$ and $p_{c}(1|0) \longrightarrow p_{c}(1|1)=1/2$ for high levels of noise.  This is in agreement with what we expect, since the background is completely unpolarized.
\par
The quantum scenario is shown in Fig. \ref{experimental_probs} b). 
We see that $p_{q}(0|0) \approx 1$ and $p_{q}(1|0) \approx 0$ for low noise.  
In the quantum case both the background noise and photon A are unpolarized.  Thus, in the absence of the object we expect $p_{q}(0|1) \approx \frac{1}{4}$ and $p_{q}(1|1) \approx  \frac{3}{4}$, as confirmed by the experimental results.   However, while the probabilities $p_c(r|0)$ for the classical strategy are still quite far from $p_c(r|1)$ for high noise levels, the probabilities $p_{q}(r|0)$ and $p_{q}(r|1)$ for the quantum strategy are visually indistinguishable for the same level of noise. This confirms the discussion following Eqs. \eqref{eq:ops}:  contrary to what is expected from the usual QTD protocol, if the detection scheme is limited to linear optics devices, the classical strategy using a separable state is more robust at high levels of noise than the quantum strategy.  In the next section we develop a simple theoretical model for the experimental probabilities presented in Fig. \ref{experimental_probs}, as a function of the intensity of the unpolarized background light. 
  
\subsection{Quantifying the noise} 
We can quantify the amount of noise by considering the joint detection events in which two uncorrelated photons, such as one photon A together with a noise photon, or two noise photons, are detected within the coincidence time window. The number of coincidence counts registered by two detectors $D_A$ and $D_B$ measuring uncorrelated sources is given by \cite{mandel95}
\begin{equation}
	NC = c_A c_B \Delta T,
	\label{eq:acc}
\end{equation}
where $c_A$ ($c_B$) is the counting rate at detector $D_A$ ($D_B$) and $\Delta T$ is the coincidence time window.  These count rates are equal to the overall detection efficiency $\epsilon$ times the number of photons incident on the detection system.  Here the notation ``NC'' stands for ``noise coincidence counts". We note that all quantities in this estimate are experimentally accessible, as it involves the count rates at each detector, and makes no assumption concerning the origin of the detected photons.  
\par
Let us consider that in a given measurement window there are $S_A$ signal photons $A$, $S_B$ signal photons $B$, and $N$ photons from the background noise, incident on our measurement device. In the classical strategy, a polarizer is used to project onto the initial separable polarization state, which at the same time removes half of the background noise. Then, using Eq. \eqref{eq:acc}, the number of coincidence counts due to noise is
\begin{align}
	NC_{c} & = \epsilon_A \epsilon_B S_A\left(S_B+\frac{N}{2}\right)\Delta T \nonumber \\ 
	& = \epsilon_A \epsilon_B S_A S_B\left(1+\frac{g}{2}\right)\Delta T , 
	\label{eq:localfilt}
\end{align}
where $\epsilon_A$ and $\epsilon_B$ are the overall detection efficiencies of detectors $D_A$ and $D_B$, respectively.   One can see in Eq. \eqref{eq:localfilt} that there are contributions to the coincidence counts that originate from joint detections of uncorrelated idler/signal events (the $S_AS_B$ contribution), as well as uncorrelated idler/noise events (the $S_A S_B g$ term).  

Let us consider now the quantum strategy consisting of the Bell-state projection. Since the central PBS combines the signal and idler modes, and each photon alone, regardless of it's origin, is completely unpolarized, the average number of photons at each output of the central PBS is proportional to $(S_A+S_B+N)/2$. The state $\ket{\phi^+}$ is then identified through an additional polarization measurement at the detectors, projecting onto $\ket{+}\ket{+}$ and $\ket{-}\ket{-}$. This gives an additional polarization filtering of the noise by a factor $1/2$, since only half of the possible unpolarized events contribute. In principle, the number of noise counts is
\begin{align}
	\label{eq:bellfilt}
	NC_{q} & = \frac{\epsilon_A \epsilon_B}{8}\left(S_A+S_B+N\right)^2\Delta T \nonumber \\
	& =  \frac{\epsilon_A \epsilon_B}{8}\left(S_A+S_B(1+g\right))^2\Delta T. 
\end{align}

The above equations show that noisy coincidence counts can arise from the joint measurement of any combination
of photons $A$, $B$ as well as noise $N$. In particular, the joint measurement of two noise photons (the $N^2$) term appears, which is not present in the classical case. Thus, the unpolarized background light contributes to the number of noise counts quite differently in the quantum case when compared to the classical one. 

To include these noise estimates in the conditional probabilities $p_\pi(r|x)$, let us write the actual coincidence counts due to photon pairs produced by the source as $SC_{\pi \phi}$ for the classical $(\pi=c$) and quantum ($\pi=q$) cases. We can write 
\begin{equation}
	C^{(x)}_{\pi \phi} = SC^{(x)}_{\pi \phi} + NC^{(x)}_{\pi \phi}
	\label{eq:Cphinew}
\end{equation}
and
\begin{equation}
	C^{(x)}_{\pi \perp} =  NC^{(x)}_{\pi \perp}.
	\label{eq:Cperpnew}
\end{equation}
 
\begin{figure}
	\includegraphics[width=8.8cm]{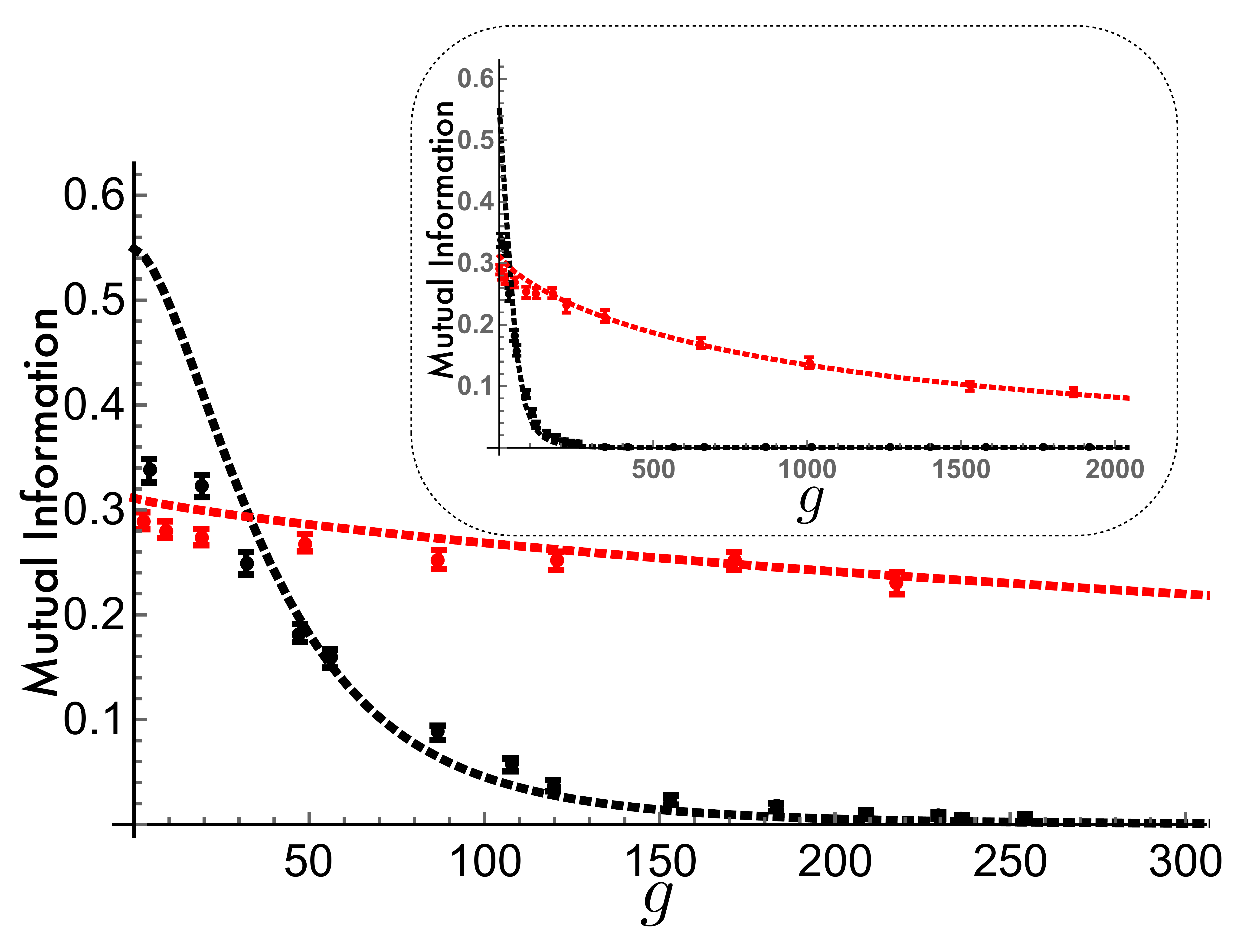}
	\caption{Experimental results for the mutual information as a function of the noise intensity $g$. Black and red lines are the theoretical predictions for the quantum and classical strategies, respectively. The dots are experimental results, which are in excellent agreement for the classical strategy. For the quantum strategy, there is a discrepancy at $g$ close to zero due to reduced two-photon interference visibility. Main figure shows a zoom of the region close to $g=0$ to show at which level of noise the classical strategy becomes better than the quantum. Inset show an evolution for larger values of noise. The parameters for our experiment are $S_A\approx 1000$, $S_B \approx 1000$, and $\Delta_T = 5$ns.}
	\label{experimental_probsb}
\end{figure}

Equations \eqref{eq:localfilt}, \eqref{eq:bellfilt}, \eqref{eq:Cphinew} and  \eqref{eq:Cperpnew} can be used in Eqs. (\ref{eq:pC}) to obtain theoretical predictions for the conditional probabilities $p_\pi(r|x)$. These are plotted in figure \ref{experimental_probs} along with the experimental data. One can see that agreement between experiment and theory is quite good, thus validating our noise model. For both strategies, the main discrepancy occurs for values of $g$ that are close to zero. In this case, the classical strategy shows a big dispersion between the experimental data of $p_c(0|1)$ and $p_c(1|1)$  and its theoretical prediction. This is caused by the small number of coincidences detected, which is originated by the lack of photons present in the absence of the object (when the noise is low).  For the quantum case, the discrepancy is  due the reduced visibility ($\sim 0.9<1$)  of the two-photon interference due to mode mismatch at the central PBS. This reduced visibility diminishes the efficiency of the Bell state measurement, achieving with probability $V$ (related to visibility) a successful Bell state measurement while with probability $(1-V)$ a noisy measurement is observed. This noise is not taken into account in our model and affects all data obtained. Thus,  when g is close to zero, all possible discrepancies between experimental and theoretical points are mainly due to this non-unit visibility. For higher values of $g$, the effect of non-unit visibility is negligible when compared to that of the background noise, observing curves with good agreement between the data and the model.

\subsection{Identifying the quantum advantage}
The capacity to identify the presence or absence of the object relies on our ability to distinguish between two probability distributions $p_\pi(r|0)$ and $p_\pi(r|1)$, corresponding to whether the object is present or absent. 
In order to identify the amount of information obtained by the measurement, we employ the mutual information between the stochastic variables $r$ and $x$, defined by $I(r:x) = H(x) - H(x|r)$ \cite{Cover}. In this equation, $H(x)$ is the binary Shannon entropy ($H(x)= -x \log x - (1-x)\log(1-x)$) while $H(x|r)$ stands for the  conditional entropy $H(x|r) = - \sum_{x,r} p(x,r) \log p(x|r)$. Such quantities can be calculated directly from experimental results by means of Bayes' rule. Assuming that the target has a probability of $1/2$ of being present or not, we have $H(x)=1$. 

Figure \ref{experimental_probsb} shows the mutual information for both the classical (red circles) and quantum (black hollow squares) cases.  Points correspond to experimental data, while the lines correspond to theoretical predictions using the noise model developed in the last section. We can see that the quantum strategy using linear optics (solid black curve) is better than the classical one (solid red curve) up to a noise/signal ratio of about $g \approx 40$. For larger values of $g$, the classical strategy presents a better performance.  As was analized previously, for values of g close to zero, the theoretical model and experimental data are not in good agreement for both  the classical and quantum strategy. The former shows a discrepancy that is related with the  large dispersion between the experimental data of $p_c(0|1)$ and $p_c(1|1)$  and its theoretical prediction. For the latter,  it is due to the non-unit visibility of the two-photon interference. 

\subsection{Improving the quantum advantage}
A simple way to improve our results is to employ coincidence electronics with a smaller coincidence window, which reduces the number of noise coincidence counts in Eqs. \eqref{eq:localfilt} and \eqref{eq:bellfilt}.  Since the temporal correlation between the source photons is extremely high (typically better than 1ps), this should not reduce the rate of signal counts.  The limiting factor here is then the temporal resolution of the detectors, which is determined by the temporal jitter ($\sim 50$ ps).  We calculated the mutual informations for the separable and entangled strategies, and found that a coincidence window of $\Delta t = 100$ps in our experiment would correspond to a quantum advantage when the signal to noise ratio $\gtrsim 1/400$.  For higher intensity of unpolarized background radiation, a non-linear optical medium can be used to perform a Bell-state measurement \cite{kim01}, though this method is highly probabilistic.    

\section{Conclusions} 
We investigated a protocol for quantum target detection based on polarization-entangled photons and a linear-optics based measurement system. The target object was immersed in unpolarized background light. The entangled probe state is distinguished from the unpolarized background noise via a partial Bell-state projector, constructed using two-photon interference and linear-optics devices.  We compare this linear-optics based strategy with a classical strategy using local projective measurements on a separable probe state. Our results shows that the linear-optics protocol allows for quantum mechanics to outperform classical strategies when the signal to noise ratio is better than about 1/40.  For higher levels of noise, the classical strategy outperforms the quantum one.  We explain this by analyzing the number of coincidence counts due to uncorrelated photons, obtaining a good agreement between theory and experiment.  These uncorrelated events originate from joint measurement of an unpolarized background photon with a signal photon, and also between two background photons.   
\par
We note that the quantum advantage we consider here concerns only the polarization correlations of the photons.  In both of our strategies, two-photon coincidence detection is employed.  Thus, even in the case of our ``classical" strategy using photon pairs in a separable polarization state, the temporal correlations of the photons provide a gain in the signal to noise ratio when compared to the use of classical light \cite{England18}.  The polarization-based quantum gain should appear in addition to the temporal-based one. 
\par
It is interesting to contrast our protocol to that of the original proposal for quantum illumination \cite{lloyd08}.  In this case, it was assumed that one can perform the optimal joint measurement, corresponding to a perfect projective measurement that distinguishes the initial state from all orthogonal subspaces.  For polarization entangled photons, one can distinguish a single Bell state ($\ket{\phi^+}_{AB}$ in the present context) from the other three Bell states using only linear optics.  These four states form a complete basis for the polarization degree of freedom when restricted to the subspace where one photon is in mode A and one in mode B ($\equiv \ket{1}_A\ket{1}_B$).  However, when the background noise is larger, it can produce non-negligible contributions of the form $\ket{0}_A\ket{2}_B$, where two photons can be found in mode B, as well as higher-order contributions consisting of more background photons.  The linear-optics Bell-state analyzer does not separate the state $\ket{\phi^+}_{AB}$ from these events perfectly.  Thus, when the amount of background noise is large compared to the number of actual signal photons, the linear-optics scheme no longer performs the optimal measurement.   We note that the optimal measurement could indeed be realized by employing a Bell-state projector based on non-linear optics \cite{kim01}. However, whereas the linear-optics device is deterministic, the non-linear device suffers from a very small success probability.  Our results should be useful in future designs of quantum sensing devices.
 
\begin{acknowledgments}
The authors would like to thank the brazilian agencies CAPES, CNPQ, FAPERJ, and the INCT-IQ for partial financial support. This work was realized as part of the CAPES/PROCAD program. 
 \end{acknowledgments}

\end{document}